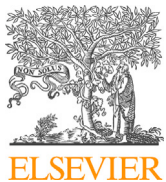
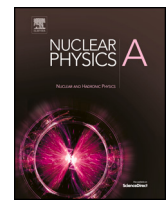

# The challenging first direct measurement of the 65 keV resonance strength in the $^{17}$O(p,$\gamma$)$^{18}$F reaction at LUNA

D. Piatti for LUNA collaboration

*University and INFN of Padova, Via Marzolo 8, Padova, 35135, Italy*

## ARTICLE INFO

## ABSTRACT

A precise determination of the proton capture rates on oxygen is mandatory to predict the abundance ratios of oxygen isotopes in a stellar environment where the hydrogen burning is active. The $^{17}$O(p,$\gamma$)$^{18}$F reaction, in particular, plays a crucial role in AGB nucleosynthesis as well as in explosive hydrogen burning occurring in novae and type I supernovae. At the temperature of interest for the former scenario (20 MK ≤ T ≤ 80 MK) the main contribution to the astrophysical reaction rate comes from the $E_r$ = 64.5 keV resonance. The strength of this resonance is presently determined only through indirect measurements, with an adopted value $\omega\gamma = (1.6 \pm 0.3) \times 10^{-11}$ eV. A new high sensitivity setup has been installed at LUNA, located at Laboratori Nazionali del Gran Sasso. The underground location of LUNA 400kV guarantees a reduction of the cosmic ray background by several orders of magnitude. The residual background was further reduced by a dedicated shielding. On the other hand the $4\pi$-BGO detector efficiency was optimized installing an aluminum target chamber and holder. With about 400 C accumulated on Ta$_2$O$_5$ targets, with nominal $^{17}$O enrichment of 90%, the LUNA collaboration has performed the first ever direct measurement of the 64.5 keV resonance strength.

## 1. Introduction

The H burning via CNO cycles, $T$ = 0.02 - 0.1 GK [33], powers massive stars, during the main sequence phase, and Red Giant Branch (RGB) and Asymptotic Giant Branch (AGB) stars [44]. Moreover, the CNO cycles are active in explosive H-burning, $T$ = 0.1 - 0.4 GK, of classical novae and type I supernovae [33].

The $^{17}$O(p,$\gamma$)$^{18}$F reaction ($Q$-value = 5607.1(5) keV [52]) fuels the third CNO cycle [45,19]:

$$^{17}O(p,\gamma)^{18}F(\beta^+\nu)^{18}O(p,\alpha)^{15}N \rightarrow$$
$$\rightarrow ^{15}N(p,\gamma)^{16}O(p,\gamma)^{17}F(\beta^+\nu)^{17}O. \qquad (1)$$

The CNOIII reaction flow is, indeed, determined by the $^{17}$O(p,$\alpha$)$^{14}$N/$^{17}$O(p,$\gamma$)$^{18}$F ratio, being the (p,$\alpha$) channel the turning point of CNOII.

In turn the $^{18}$O(p,$\gamma$)$^{19}$F reaction, with $^{18}$O being the daughter of the radioactive $^{18}$F ($T_{1/2}$ = 109.77(5) min [51]), may trigger a fourth cycle. The CNOIV was recently suggested to pave the way to the NeNa cycle through the $^{19}$F(p,$\gamma$)$^{20}$Ne reaction, fueling the calcium overproduction in population III stars [53,18,54].

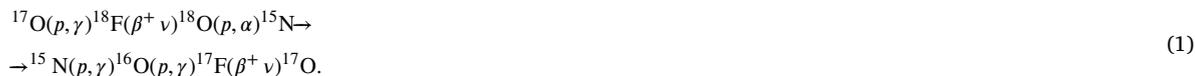

In RGB and AGB stars the ashes of H-shell burning, enriched in $^{17}$O, are mixed with the stellar surface material during several mixing processes, causing a change of the surface $^{16}$O/$^{17}$O ratio, which can be observed from the radio and infrared spectra of giant stars [31,32,35] and in presolar grains [57]. Once the cross section of CNO reactions are precisely determined the comparison between the model predictions and the observed oxygen isotopic ratio provides a powerful tool to constrain both the depth attained by the mixing episodes [50,32,36] and their origin, such as convection, rotation, magnetic buoyancy, gravity waves, and thermohaline circulation [20,8,24,17,39,21,14,23,40]. Moreover a precise knowledge of the $^{17}$O+p reaction cross sections would help in unveiling the nature of the progenitors of the stardust grains [37].

Ultimately the $^{16}$O/$^{17}$O ratio is used to trace the Galactic Chemical Evolution. The long standing puzzle of the $^{17}$O overproduction predicted by Galactic Chemical Evolution models [42,46,50] was only mitigated by a recent direct measurement of the $E_r$[1] = 64.5 keV resonance strength in $^{17}$O(p,$\alpha$)$^{14}$N channel [11].

On the other hand the 64.5 keV resonance in the (p,$\gamma$) channel is poorly constrained.

The strength of the $E_r$ = 64.5 keV resonance is presently determined only through indirect measurements, the $\Gamma_\gamma$ and $\Gamma_\alpha$ were provided by measurement of the $^{14}$N($\alpha,\gamma$)$^{18}$F and $^{14}$N($\alpha,\alpha$)$^{14}$N reaction respectively [4,38]. The $\Gamma_p$ is derived from the $\omega\gamma$ of the $^{17}$O(p,$\alpha$)$^{14}$N channel and it contributes the most to the final uncertainty because of the discrepant results reported in literature [10,13,47]. The presently adopted resonance strength is $\omega\gamma_{(p,\gamma)} = (16 \pm 3)$ peV [13]. Such a low resonance strength translates in an expected rate as low as one reaction per Coulomb, thus a direct measurement of the $E_r$ = 64.5 keV resonance strength required both a high sensitivity setup and a devoted technique to monitor and subtract potential beam-induced background (BIB). In the following sections the setup used, the analysis performed and the results obtained by LUNA collaboration in the first direct measurement of $E_r$ = 64.5 keV resonance strength [29] will be reported.

## 2. Experimental setup

The measurement was performed at LUNA laboratory, located in the deep underground facility of Laboratori Nazionali del Gran Sasso (LNGS) [16]. Thanks to the 1400 m overburden of rock, the muon cosmic ray background is reduced by six orders of magnitude with respect to the surface. The residual background at $E_\gamma \leq 3$ MeV is due to the natural radioactivity from the laboratory and the rock while at higher energy it is mainly due to the neutron-induced background [9]. In order to increase our sensitivity, a dedicated setup was designed to maximize the detection efficiency and to reduce the residual background, detailed description of the upgrades and the achieved sensitivity can be found in [49]. For the sake of clarity we will report here the setup main features only.

The high stability and intensity (200 $\mu$A) proton beam provided by LUNA400kV accelerator [26] was delivered through a Cu pipe to the target. A negative voltage was applied to the copper tube to act as secondary electron suppressor and it was in thermal contact with liquid nitrogen to act as cold finger. The Ta$_2$O$_5$ solid targets, of two different thicknesses 147 and 378 nm, were produced by anodization of tantalum backings in 90% $^{17}$O enriched water doped with 4% $^{18}$O [15]. The target was water cooled to prevent degradation, which was monitored via periodical scan of the $E_r$ = 143 keV resonance in the $^{18}$O(p,$\gamma$)$^{19}$F reaction and with periodic run on top of the $E_r$ = 183 keV resonance in the $^{17}$O(p,$\gamma$)$^{18}$F reaction [22]. To monitor the beam induced background, additional targets were produced, performing the anodic oxydation in a solution of Ultra Pure Water (UPW). The UPW-targets had the same thicknesses as oxygen-17 targets but negligible amount of $^{17}$O isotope.

In order to minimize the $\gamma$-ray absorption, both the scattering chamber and the target holder were made in aluminum, providing an increase in efficiency of at least 18% over the whole energy range with respect to the previous stainless-steel and brass setup [49,29]. The LUNA Bismuth-Germanium-Oxide (BGO) detector surrounded the reaction chamber, covering a 4$\pi$ angle. The detector is made of six optically independent crystals, which coupled with a listmode DAQ allow both a single crystal reading and the construction of the add-back spectrum, namely by adding coincident events in the individual crystals [6]. The residual background was further reduced by a three layer shielding which was installed all around the detector and the target chamber. The shielding is made of 1 cm thick layer of borated (5%) polyethylene, 15 cm thick lead shielding and 5 cm thick borated (5%) polyethylene envelope. The detected background was reduced by a factor 4.3 $\pm$ 0.1 in the region of interest (ROI) for our measurement (5.2 - 6.2 MeV), with respect to using only lead [7,49,29]. A 3D model of the detection setup is shown in Fig. 1.

## 3. Analysis and results

The data taking covered 4 months, with an overall accumulated charge of 420 C and 300 C on oxygen-17 and UPW-targets, respectively. Long runs (~12 hours each) have been performed at $E_p$ = 78 keV to populate the 64.5 keV resonance well inside the target. Scans of the 143 keV resonance in the $^{18}$O(p,$\gamma$)$^{19}$F reaction and runs on top of the 183 keV resonance in the $^{17}$O(p,$\gamma$)$^{18}$F reaction alternated the long runs to monitor the target degradation.

The 64.5 keV resonance strength was determined using a thick-target approach [34]:

$$\omega\gamma = \frac{2}{\lambda^2}\epsilon_{eff}\frac{N_\gamma}{\eta W N_p}, \qquad (2)$$

where the $\lambda$ is the de Broglie wavelength at the center-of-mass resonant energy. The number of incident protons, $N_p$, was derived, with a 2% uncertainty, integrating the beam current directly from the target chamber, which acted as a Faraday Cup [11]. The angular distribution, $W$, is 1 since the BGO covers a $4\pi$ solid angle. The effective stopping power, $\epsilon_{eff}$, in the center-of-mass frame, was calculated using SRIM database [56] with a 4% total uncertainty. The $\gamma$-ray detection efficiency, $\eta$, was measured up to 2 MeV using point-like radioactive sources ($^{137}$Cs, $^{60}$Co and $^{88}$Y), with activities calibrated by the Physikalisch-Technische Bundesanstalt (PTB) to 1% accuracy. To cover the region of interest of the present investigation, the efficiency data were extended up to 7.6 MeV using the well-known $^{14}$N(p,$\gamma$)$^{15}$O resonance at proton energy $E_p$ = 278 keV [27,30]. A Geant4 [1] based simulation of the present setup, validated against the aforementioned calibration measurements, allowed to fully characterize the BGO response within 3% uncertainty, more details in [49].

---

[1] In the following we will indicate with $E$ or either $E_r$ the proton energy in the center of mass, while with $E_p$ the proton energy in the laboratory system.

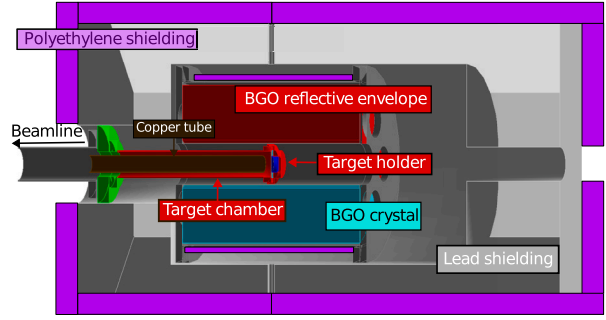

**Fig. 1.** Sketch of the setup as from GEANT4 simulation code [49].

To calculate the net counts, $N_\gamma$, from $^{17}$O(p,$\gamma$)$^{18}$F reaction, first the long runs were precisely calibrated up to 8 MeV using the 143 keV scan runs, performed immediately before and after. The region of interest (ROI) for the addback spectrum was determined via both a dedicated study of BGO resolution and via simulation of the $E_x$ = 5672 keV de-excitation cascades. Then to calculate the net counts the contribution of both environmental and beam induced background, and of the direct capture component must be subtracted. The background was dominated by the single $\gamma$-rays from the $^2$H(p,$\gamma$)$^3$He (Q-value = 5493.47508(6) keV [52]) beam induced reaction, with deuterium stored in Ta backings [2]. Due to the poor BGO energy resolution the 64.5 keV resonance sum peak and the $^2$H(p,$\gamma$)$^3$He direct capture peak overlap, hampering a precise discrimination. To distinguish the $\gamma$-rays of interest from those emitted by $^2$H(p,$\gamma$)$^3$He reaction we exploited both the BGO segmentation and the well known cascades of the $E_x$ = 5672 keV level [51]. The so called gate analysis, detailed in [49,29], has been proved to be a successful tool [25,41,55] and for the present case it was implemented selecting events contributing to the sum peak and matching the condition of multiplicity > 1 and with $\gamma$-ray deposited energy in single crystal corresponding to $3300 \leq E_\gamma \leq 4850$ keV range. The latter condition allows to select the primary $\gamma$-rays of the transitions to the 1080 and 1041 keV states in the de-excitation scheme of the $E_x$ = 5672 keV level, see Fig. 2. These filters applied to the sum peak events allowed to fully discriminate between the 64.5 keV resonance contribution and the $^2$H(p,$\gamma$)$^3$He reaction, which proceeds, indeed, to the ground state via a single $\gamma$-ray emission. The same discrimination analysis as on the experimental spectra were applied to simulation of the 64.5 keV resonance in order to obtain the proper detection efficiency. A 7% total uncertainty on the efficiency was estimated taking into account the branching ratio uncertainty (6%) and the simulation accuracy (3%).

Random coincidences from beam induced and natural background, mimicking the case of interest are still possible. To properly subtract these counts the same discrimination procedure was applied to long runs performed with UPW-targets. After a charge based normalization, the random coincidences were subtracted [29]. Finally, to calculate the direct capture contribute to the observed count rate, $Y_{DC}$, the following relation was used:

$$Y_{DC} = \eta_{DC} \int_{E_p - \Delta E_p}^{E_p} E^{-1} S(E) e^{-2\pi\eta} \epsilon_{\text{eff}}^{-1}(E) P(E) dE$$

where $E_p$ is the proton beam energy, $\Delta E_p$ the target thickness in terms of energy lost by the beam, $\epsilon_{\text{eff}}(E)$ the effective stopping power, and $e^{-2\pi\eta}$ the Gamow factor [44]. The S-factor, $S(E)$, was extrapolated via R-matrix fit of the data available in literature, details in [43]. The branching ratios were implemented in the simulation code as reported in literature [22]. The gate analysis was applied to simulations, providing the efficiency, $\eta_{DC}$, for the DC component mimicking the signal of interest. The DC contribution was 8% of the measured experimental yield [29].

The present result, $\omega\gamma_{(p,\gamma)} = (34 \pm 7_{\text{stat}} \pm 3_{\text{syst}})$ peV was corrected for the electron screening, correction factor $f = 1.15$ derived considering the adiabatic approximation [3] resulting in a $\omega\gamma_{(p,\gamma)}^{\text{bare}} = (30 \pm 6_{\text{stat}} \pm 2_{\text{syst}})$ peV [29].

The present result is the first obtained by a direct measurement and it is higher than values reported in literature by a factor ~2. The proton width calculated using present result is $\Gamma_p = (39 \pm 8_{\text{stat}} \pm 3_{\text{syst}})$ neV (corresponding to $\Gamma_p^{\text{bare}} = (34 \pm 7_{\text{stat}} \pm 3_{\text{syst}})$ neV) [29], in good agreement with result by [11].

## 4. Summary and conclusions

In summary, we reported on the first direct measurement of the 64.5 keV resonance strength in $^{17}$O(p,$\gamma$)$^{18}$F reaction. The deep underground location of LUNA, the improvements to the setup and a the application of the gate analysis allowed to achieve the desired sensitivity. Our result for the resonance strength is about a factor of 2 higher than previous values reported by [5,12,48,34,28], suggesting a higher $\Gamma_p$ in agreement with previous result by LUNA on 64.5 keV resonance in (p,$\alpha$) channel [11]. Detailed astrophysical implications of the new suggested rate for the (p,$\gamma$) channel will be reported in a forthcoming work.

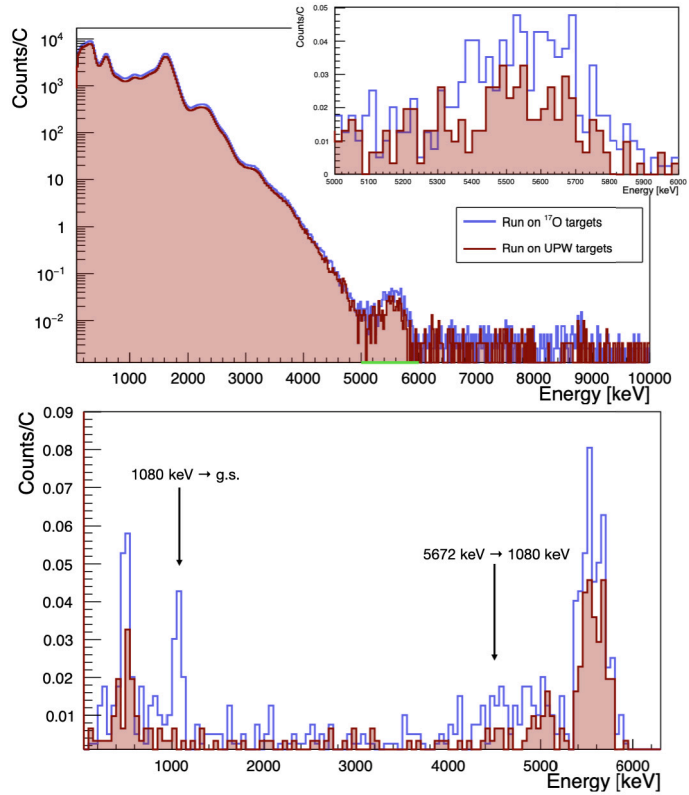

**Fig. 2.** Gate analysis applied on total statistics accumulated on top of the 64.5 keV resonance with both oxygen-17 targets, in red, and UPW targets, in blue [29]. (For interpretation of the colors in the figure(s), the reader is referred to the web version of this article.)

**Declaration of competing interest**

The authors declare the following financial interests/personal relationships which may be considered as potential competing interests: Denise Piatti reports financial support was provided by Università e INFN di Padova. Denise Piatti reports a relationship with University of Padua that includes: employment. Denise Piatti has patent licensed to LUNA collaboration. None If there are other authors, they declare that they have no known competing financial interests or personal relationships that could have appeared to influence the work reported in this paper.

**Acknowledgements**

D. Ciccotti and the technical staff of the LNGS are gratefully acknowledged for their help during setup construction and data taking. Dr. Sara Carturan from LNL and the chemistry laboratory staff of LNGS are acknowledged for the help with target preparation. Financial support by INFN, the Italian Ministry of Education, University and Research (MIUR) (PRIN2022 CaBS, CUP:E53D230023 and SOCIAL, CUP:I53D23000840006) and through the "Dipartimenti di eccellenza" project "Science of the Universe", the European Union (ERC Consolidator Grant project *STARKEY*, no. 615604, ERC-StG SHADES, no. 852016), (ELDAR UKRI ERC StG (EP/X019381/1)) and (ChETEC-INFRA, no. 101008324), Deutsche Forschungsgemeinschaft (DFG, BE 4100-4/1), the Helmholtz Association (ERC-RA- 0016), the Hungarian National Research, Development and Innovation Office (NKFIH K134197), the European Collaboration for Science and Technology (COST Action ChETEC, CA16117) and the Hungarian Academy of Sciences via the Lendület Program LP2023-10. C. G. B., T. C., T. D. and M. A. acknowledge funding by STFC UK (grant no. ST/L005824/1).

**Data availability**

Data are already published